\documentstyle[12pt,aasms4]{article}
\def\ni{\noindent}
\def\spose#1{\hbox to 0pt{#1\hss}}
\def\lta{\mathrel{\spose{\lower 3pt\hbox{$\mathchar"218$}}
     \raise 2.0pt\hbox{$\mathchar"13C$}}}
\def\gta{\mathrel{\spose{\lower 3pt\hbox{$\mathchar"218$}}
     \raise 2.0pt\hbox{$\mathchar"13E$}}}

\begin{document}

\title{Energetics and Luminosity Function of Gamma-ray Bursts}
\author{Pawan Kumar}
\affil{JILA, Boulder, CO 80303, and IAS, Princeton, NJ 08540}
\author{Tsvi Piran}
\affil{Racah Institute, The Hebrew University, Jerusalem 91904, Israel}
\authoremail{pk@ias.edu,tsvi@nikki.fiz.huji.ac.il}

\begin{abstract}
\baselineskip 15pt

Gamma-ray bursts are believed to be some catastrophic event in which
material is ejected at a relativistic velocity, and internal
collisions within this ejecta produce the observed $\gamma$-ray flash.
The angular size of a causally connected region within a relativistic
flow is of the order the angular width of the relativistic beaming,
$\gamma^{-1}$. Thus, different observers along different lines of
sights could see drastically different fluxes from the same
burst. Specifically, we propose that the most energetic bursts
correspond to exceptionally bright spots along the line of sight
on colliding shells, and do not represent much larger energy
release in the explosion. We calculate the distribution function of
the observed fluence for random angular-fluctuation of ejecta.  We
find that the width of the distribution function for the observed
fluence is about two orders of magnitude if the number of shells
ejected along different lines of sight is ten or less. The
distribution function becomes narrower if number of shells along
typical lines of sight increases. The analysis of
the $\gamma$-ray fluence and afterglow emissions for GRBs with known
redshifts provides support for our model i.e. the large width of 
GRB luminosity function is not due to a large spread in the energy 
release but instead is due to large angular fluctuations in ejected 
material. We outline several observational
tests of this model. In particular, we predict little correlation
between the $\gamma$-ray fluence and the afterglow emission as in fact is
observed. We predict that the early (minutes to hours)
afterglow would depict large 
temporal fluctuations whose amplitude decreases with time.
Finally we predict that there should be many weak bursts with
about average afterglow luminosity in this scenario. 

\end{abstract}

\bigskip
\hskip0.3cm{\it Subject headings:\rm~ gamma rays: bursts -- relativistic shock}

\vfill\eject
\baselineskip 15pt

\section{Introduction}

The improvement in the determination of angular position of GRBs by
the Dutch--Italian satellite, BeppoSAX, has led to measurement of
distances to eight GRBs and has lent support for the relativistic shock 
model (cf. Costa et al. 1997, van Paradijs et al. 1997, Bond 1997, Frail et
al. 1997).  Unlike previous expectations these observations revealed
that the GRB luminosity function  is  very broad; the width of the 
fluence distribution is about two orders of magnitude in energy.
Assuming isotropic emission one finds that the energy of the most energetic
bursts is larger than $10^{54}$ergs. The recently observed evidence 
for beaming with an opening angle of
a few degrees reduces the energy estimate by a factor of a hundred
(Kulkarni et al., 1999; Sari, Piran \& Halpren, 1999; Harrison et
al., 1999).  However, when taking into consideration the relatively low
efficiency of conversion of kinetic energy to $\gamma$-rays, one finds
that  the total
kinetic energy required is, after all, of order 10$^{54}$ erg even
after the beaming correction (Kumar, 1999). This energy is too
large to be released by a compact solar mass object. According to the
internal-external shock model comparable amount of energy should be
released during the GRB phase and the afterglow. However, quite generally,
only a fraction of the energy emitted as $\gamma$-rays is seen in the 
afterglow (mostly as X-rays).
Moreover, there appears to be little correlation between the
$\gamma$-ray fluence and the afterglow flux.

All the above mentioned phenomena could be unrelated. However, within
the framework of relativistic internal shocks model  
(Narayan, Paczynski \& Piran, 1992, Paczynski \& Xu 1994,
Rees \& M\'esz\'aros 1994, Sari \& Piran 1997) we suggest
a possible connection: these properties could all be manifestations of
large angular inhomogeneity of the relativistic ejecta. The size of a 
causally connected region of a shell of radius $R$ moving with Lorentz 
factor $\gamma$ is $\sim R/\gamma$. Because of relativistic
beaming this is also the size of  the region visible to a
distant observer. During the GRB the angular size of these regions
($<0.01$ rad) is significantly smaller than the inferred angular 
width of the ejecta, $\delta \theta\sim$ a few degrees. There are 
therefore $(\gamma\delta\theta)^2$ causally disconnected regions
within this cone. Thus the observed $\gamma$-ray luminosity
seen by different observes from the same burst could fluctuate
strongly due  to small scale inhomogeneities in the emitting
regions. Unless careful, one would over-estimate the energy release
in $\gamma$-rays in cases in which a hot spot has been 
observed\footnote{Note that as BeppoSAX can detect only rather strong 
bursts the BeppoSAX sample might be biased towards cases in which such 
a hot spot has been seen.}. At a later time when the Lorentz factor 
of the ejecta has become smaller and the size of causally connected 
regions is larger the dispersion of the afterglow luminosity seen 
along different lines of sight should be smaller as well. The emission
at this stage yields a better estimate of the overall total energy involved.

We explore, here, the implications of this model. The model is
presented in the next section, followed by calculation of the
distribution of fluence. The observational aspects are discussed in
section 3, and the main results and some predictions of the model
are summarized in section 4.

\section{The Physical Model and Numerical simulation}
\label{physical}

The calculations described in this section are numerical, 
but in \S2.2 we analyze a simple model analytically to gain 
some insight into the numerical results.

We consider successive, random, ejection of blobs of angular size
$\gamma^{-1}$ into a cone with an opening angle $\delta\theta = 10^o$. 
To estimate the expected fluence distribution we
have carried out Monte Carlo simulations of the observed emission from
randomly ejected shells - corresponding to the random conditions
expected in different causally disconnected regions.  
In each case we keep the overall energy ejected into a given
cone of opening angle 10$^o$ approximately the same -- $10^{52}$ergs. 
Changing the total energy in bursts causes a linear translation of 
the fluence distribution function.
The Lorentz factor of each shell is assumed to be a random number 
uniformly distributed between $\gamma_{min}$ and $\gamma_{max}$; 
we take $\gamma_{min}=5$ \& $\gamma_{max}=400$. The
energy distribution of blobs is taken to be either log-normal with mean
of 10$^{52}/N_b$ erg and width (FWHM) in log$_{10}$(energy) of 1, or
a delta function distribution;
$N_b$ is the number of blobs ejected in the explosion. 
The number of blobs ejected along a line of sight is also taken to be a 
random number uniformly distributed between $N_{min}$ and $N_{max}$.  
We have considered two different values for the average number
of shells ejected: $(N_{min} + N_{max})/2$ equal to 5 and 40.
The shells are ejected randomly over some time interval corresponding 
to the total duration of  30 s.

The calculation and the results described here apply to the case 
where shells have little or no angular fluctuation, and the radiation
received by different observers located within 10$^o$ cone
is a smooth function. The distribution function for fluence 
in this case is significantly narrower compared to when shells 
have large angular fluctuations (see \S2.1).

We consider all possible shell collisions along a line of sight, and
for each collision we follow the forward and reverse shocks
propagating within the colliding shells. The thermodynamic properties
of the shocked gas are calculated by solving the continuity of energy,
momentum, and baryon number flux  (see e.g. Piran 1999).
Shell collisions continue until mergers arrange
the shell velocities as a monotonically increasing function of their
distance from the center of explosion. 
We assume equipartition of energy between electrons, protons and 
magnetic field (the energy fraction in magnetic field $\epsilon_B=0.3$, 
and in the electrons $\epsilon_e=0.3$). The distribution of electron
number density is taken to be a power-law function of energy, $n_e(E)\propto
E^{-p}$, with $p=2.5$. 

We calculate synchrotron emission from relativistic electrons and
include the effect of electron cooling and inverse-synchrotron absorption
on the power spectrum (e.g. Sari, Piran \& Narayan, 1998).
The synchrotron photons undergo inverse Compton scattering to
produce the emergent spectrum (we calculate multiple Compton scatterings
when the Compton $y$-parameter is greater than 1). As photons emitted 
by one shell might undergo elastic collisions in other shells we follow 
the trajectory of photons produced in shell collisions, along with the 
trajectory of different shells, to determine the energy and momentum 
deposited by photons in different shells and the resulting change to 
the bulk kinetic energy of shells (Kumar, 1999).

The computed emergent power spectrum is integrated between 10--10$^3$ keV, 
in the observer frame, to determine fluence along different lines of
sight. From the `observed' fluence we calculate the isotropic
radiative energy in $\gamma$-rays (it should be emphasized, however,
that the radiation is highly anisotropic when shells are not uniform). 
The total energy in explosion is obtained by adding the energy of all 
the blobs in a cone of 10$^o$ opening angle, which as stated previously is 
taken to be 10$^{52}$ erg.

\subsection{Fluence distribution and Burst Energetics}
\medskip

Figure 1 shows the isotropic $\gamma$-ray fluence distribution resulting from
the calculation described above where the total energy in the explosion
is 10$^{52}$ erg. The full width at half maximum (FWHM) of 
log$_{10}$(fluence) distribution function is about 2.0, i.e. two orders of 
magnitude in the linear energy scale, 
when the mean number of shells along the line of sight is 5 and the
FWHM of the energy distribution of individual blobs is 1.0 (on log$_{10}$
scale). Part of the width of the fluence distribution results from the 
energy distribution of blobs, and rest from the distribution of 
their Lorentz factor. The FWHM of the fluence distribution function
is 1.2, i.e. a factor of 15 on linear scale, when the energy of 
blobs follows a delta-function distribution. Clearly in this case fluence 
distribution is a result of the fluctuation in the radiative efficiency
of internal shocks.

The width of the fluence distribution function decreases with
increasing number of blobs along the line of sight; the FWHM
is only half an order of magnitude when the mean number of shells
along a direction is 40 (fig. 1). The width of the distribution
function is not sensitive to burst duration, and moreover it is
almost independent of the energy fraction in magnetic field as 
long as $\epsilon_B$ is not so small that the shocks
become almost adiabatic.  It is also evident from fig. 1 that there
are a number of bright spots in the cone containing the ejecta for
which the observed isotropic $\gamma$-ray fluence is 5x10$^{53}$ erg.

The fluence distribution function for the model consisting of collisions
of uniform shells is similar to the graph in fig. 1. There are two 
differences, however, between the uniform and the patchy shell models.
One, the width of the fluence distribution function is smaller
by a factor of $\sim 10$ for the uniform shell model with the same
fixed total energy in explosion. The other difference is that
the afterglow flux falls off smoothly in this model whereas in the 
case of inhomogeneous shells the afterglow flux at early time, within the
first hour of the explosion, has large amplitude fluctuation. 
As discussed in \S4 this can be used to observationally 
distinguish between the two models.

We have assumed that consecutive ejection of shells is completely 
random. It is possible that the energies of blobs ejected along a 
given line sight might have a non-zero correlation.
It is straightforward to modify our calculation to include this
correlation. We find that even for a perfect correlation of blob
energy, and random distribution of Lorentz factor, the width of the 
distribution function for fluence is only slightly larger than the case
where the correlation is zero (see \S2.2 for discussion).

The integral probability distribution function is also shown in fig. 1. 
Note that the probability of seeing a burst with isotropic 
fluence of 7x10$^{53}$ erg is about 0.1\% (the total energy in explosion
is fixed at 10$^{52}$ erg). Thus, a particularly bright
burst such as GRB990123, which was in the top 0.1\% of all BATSE
bursts, could well have had total energy in explosion similar to an 
average BATSE burst. In the case of GRB990123 we were perhaps
looking at a very bright spot on the colliding shell surface.
 
This suggests that $\gamma$-ray fluence is not a reliable measure of
the total energy in explosion.
A more promising way to estimate the energetics of explosion is
to consider emission at a later time when a number of causally 
disconnected regions have merged -- but the shock is still radiative --
thereby reducing the dispersion seen along different lines of sight.
The ratio of the synchrotron cooling and the dynamical time is
$t_s/t_d \approx 6\pi m_e c/(\sigma_T t_{obs} B^2 \gamma^2)\approx
m_e/(\epsilon_B\sigma_T m_p c n_{ism} t_{obs} \gamma^4)\propto t_{obs}^{1/2}$,
which becomes greater than 1 at $t_{obs} \sim 1$ hr. At this time
$\gamma\sim 30$ and we expect about 20 disconnected regions to have
merged, and the dispersion of the radiative flux along different lines of
sights to have decreased by a factor of about 4. Note that the 
energy of the ejecta has also dropped by a factor of about 4 at this time.

\subsection{Analysis of a simple model to understand the numerical results}
\medskip

We Analyze a simple model that captures some of the features of the
numerical result presented above. Let us consider collision of two
shells. The energy and the Lorentz factor of the outer
(denoted by subscript 1) and the inner shell (subscript 2) are
given by a distribution function $P_i(E,\gamma)$. The energy radiated
when shells collide in some fraction of the total thermal energy produced 
in shell collision which is given by 
(cf. Kobayashi et al. 1998),
$$E_{col} = (E_1+E_2) \left[1 - \left({E_1\over\gamma_1} +
{E_2\over\gamma_2} \right) \left( {E_1^2\over\gamma_1^2} +
{E_2^2\over\gamma_2^2} + {2E_1 E_2\gamma_r\over
\gamma_1\gamma_2}\right)^{-1/2} \right], \eqno(1)$$ where $\gamma_r =
\gamma_1\gamma_2(1 - v_1 v_2)\approx \gamma_1\gamma_2( \gamma_1^{-2} +
\gamma_2^{-2})/2$ is the relative Lorentz factor of collision of the
shells. The thermal energy release is a function of
$\gamma_2/\gamma_1\equiv\eta$, and the energy radiated in some
frequency band is a fraction of $E_{col}$. For instance, if electrons,
magnetic field and protons are in equipartition, the energy radiated
in 10--10$^3$ keV is roughly $E_{col}/10$ (Kumar, 1999; Panaitescu et
al. 1999). To simplify the analysis we calculate the bolometric
fluence distribution function.  The distribution in some other band is
not all that different.

The bolometric distribution function is given by
$$ P_{rad}(>F) = \int_{\gamma_{min}}^{\gamma_{max}} d\gamma_1
   \int_{\gamma_1}^{\gamma_{max}} d\gamma_2 \int_{E_{1min}}^\infty dE_1
   \int_{E_{2min}}^\infty dE_2\; P_1(E_1,\gamma_1) P_2(E_2,\gamma_2), \eqno(2)
$$
where $E_{1min}\approx {F/[ (\eta+1) (\eta^{1/2} - 1)^2]}$ is the
minimum energy of the outer shell, for a given $\eta=\gamma_2/\gamma_1$, so
that the radiated energy is $F$, and $E_{2min} \approx {2F\eta^2/ (\eta-1)^2}$
is the minimum energy of the inner shell to yield fluence $F$ in shell 
collision.

Let us assume that the distribution functions are separable and write them
as $P_i(E,\gamma) = P_{iE}(E) P_{i\gamma}(\gamma)$. Substituting this 
into equation (2) we obtain

$$ P_{rad}(>F) = \int_{\gamma_{min}}^{\gamma_{max}} d\gamma_1
   \int_{\gamma_1}^{\gamma_{max}} d\gamma_2\;\; P^c_{1E}(E_{1min}) 
   P^c_{2E}(E_{2min}) P_{1\gamma}(\gamma_1) P_{2\gamma}(\gamma_2), \eqno(3)$$
where $P^c_{iE}(E) \equiv \int_E^\infty dE'\, P_{iE}(E')$.

Let us consider a particularly simple case where the distribution function
for the inner and the outer shells are both constant in the
intervals (0,$E_{max}$) and $(\gamma_{min}, \gamma_{max})$ and
zero outside. The above equation can be easily integrated in this case
and we find that the differential distribution function, $P'_{rad} \equiv
dP_{rad}/dF$, is approximately proportional to $F^{-1/2}$ for
$F\ll E_{max}/2\approx F_{max}$, becomes steeper for $F\sim F_{max}$
and is zero for $F>F_{max}$. It can also be shown that for
$P_{iE}\propto E^{\alpha}$, $P'_{rad}(F)\propto F^{-1/2}$ for
$\alpha\ge 0$, and $P'_{rad}$ falls off more steeply with $F$
for $\alpha<0$. For $P_{iE}(E)\propto \delta(E-E_0)$, 
$P'_{rad}(F)\propto F^{-1/2}$ as well i.e. the fluence distribution has
a finite width even when the energies of all the blobs are identical.

Now we turn to collision of more than 2 shells. Let us assume that collision
of any two shells gives rise to a distribution of fluence that is
same as calculated above. This is a drastic simplification,  
nevertheless it provides useful insight into the qualitative 
behavior of the fluence-distribution function calculated in \S2.2.
The distribution function resulting from $N$ collisions can be easily
calculated and is given by:

$$ P_{rad}^{(N)}(F) = {1\over (2\pi)^{1/2}} \int_{-\infty}^{\infty} dk\;
   P'_{rad}(k)\exp(-ikF) \left[ \int_0^{F} dF'\; P'_{rad}(F')\exp(ikF')
    \right]^{N-1}, \eqno(4)$$
where $P'_{rad}(k)$ is the Fourier transform of $P'_{rad}(F)$.
For $P'_{rad}\propto F^{-\alpha}$ we find from the above equation 
that $P_{rad}^{(N)}(F) \propto F^{(1+\alpha)N-1}$ for $F\ll F_{max}$, 
and it is zero for $F>N F_{max}$. The width of the distribution 
function on Log$_{10}$ scale is $3/[(1-\alpha)N + 2]$. The effective 
value of $\alpha$ is about 0.7 for the case where the energy spectrum 
of ejected shells is flat. The width calculated for the toy problem
is somewhat smaller (for fixed $N$) than for the more realistic 
problem considered above because the energy produced in shell 
mergers decreases in successive mergers, hence the effective $N$ 
for the merges is smaller than the number of shells expelled in 
the explosion.  We show the distribution function for the model 
problem in fig. 2 for $\alpha=0.5$ and $0.7$.

To determine the effect of correlated ejection of shells we set $P_2(E_2,
\gamma_2) = \delta(E_1-E_2) P_{2\gamma}(\gamma_2)$ in equation (2). We find
that for $F\ll F_{max}$, $P'_{rad}(F)$ is not very different from the case 
where shells were randomly ejected with no-correlation, whereas
for $F\sim F_{max}$, the distribution function ($P'_{rad}$) falls-off 
somewhat more rapidly, which causes $P_{rad}^{(N)}$ to become a bit broader.

\section{Comparison with observations}
\medskip

Table I depicts a compilation of the observations of GRBs and their 
afterglow with known redshifts. Included in the table is the observed 
isotropic $\gamma$-ray fluence (from Band et al., 1999), the X-ray 
luminosity after 5 hours (estimated from the published X-ray flux 
and the slope of the light curve), and the R-band magnitude 24 hours 
after the burst. To obtain a uniform sample we consider only 
bursts that have been observed by BATSE, for which there is a well 
determined fit for the spectrum using the Band function (Band et al., 1993). 
It can be seen from the table that while the spread in the isotropic 
$\gamma$-ray fluence is about two and a half orders of magnitude, the 
spread in the isotropic X-ray luminosity 5 hours after the burst, and 
the R-band optical luminosity at 1 day, is only about one and a half orders 
of magnitude. 

These dispersions can be quantified. Consider first the six bursts
with known redshifts and a well determined BATSE
fluences\footnote{Note that eight bursts appear in table I. However,
the calculation of $\sigma_\gamma$ is based on a subset of six bursts
for which there is $\gamma$-ray data.  Other subsets of six bursts are
used to calculate $\sigma_x$ and $\sigma_R$.}. Assuming a normal 
distribution  of $log(E_\gamma)$
we find that the standard deviation of the logarithm of
the (isotropic) energy emitted in $\gamma$-rays, $\sigma_\gamma$, is
0.87.  The corresponding FWHM is 2. The average isotropic $\gamma$-ray
fluence is $1.4 \times 10^{53}$ergs. The likelihood is larger than
0.05 of the maximal value within the range $0.45 < \sigma_\gamma <
2.4$. Similar analysis for the isotropic X-ray luminosity 5 hours
after the burst for six bursts yields: $\sigma_x = 0.58$, and FWHM of
1.4. The average (isotropic) X-ray luminosity is $1.3 \times
10^{46}$ergs/sec. A variance range for likelihood larger than 0.05 of
the maximal likelihood is: $0.35 < \sigma_x < 1.45 $. The standard
deviation of the logarithm of the R-band luminosity 24 hours after the
burst is $\sigma_R = 0.53$.
We note that we have not corrected the observed flux for extinction 
and thereby have overestimated $\sigma_R$. A prediction of the patchy
shell model is that $\sigma_R$ should be smaller than $\sigma_x$. Current
observations are consistent with this expectation, however more accurate
determination of $\sigma_R$ and larger number of afterglows are needed to 
improve the statistical significance of this result.

These results show that the FWHM of $\gamma$-ray energy distribution
is wider by a factor of five than the X-ray afterglow luminosity
distribution and is roughly consistent with the expected decrease in
fluctuation amplitude by a factor of 7 based on the merger of causally
disconnected regions ($\gamma$ decreases by a factor of about 7 at 5
hr). If the $\gamma$-ray emitting surface were uniform (not highly
patchy as considered here) and the large width of the isotropic
$\gamma$-ray energy distribution were due to a wide distribution of
the explosion energy (or the opening angle of the jet) then the distribution 
of afterglow luminosity in the X-ray and other wavelengths should have
been wider than the $\gamma$-ray energy distribution since the
afterglow flux $\log f_\nu = {1\over 4}(p+3)\log E + 0.5 \log n -
{3\over 4} (p-1) \log t_{obs} + constant$; where $p\approx 2.5$, $n$
is the density of the circumstellar medium, $E$ is the energy in the
explosion per unit solid angle so long as the opening angle of the
ejecta is larger than $\gamma^{-1}(t_{obs})$, and the $constant$ term
includes the dependence on $\epsilon_e$ and $\epsilon_B$. Assuming that 
$E$ and $n$ are uncorrelated, we expect the width of the afterglow luminosity,
$\log(f_\nu)$, to be larger than the width of $\log E$ distribution 
by at least a factor of 1.4. Since the observed X-ray afterglow luminosity
distribution is narrower by a factor of $\sim 1.5$ compared to
$\gamma$-ray fluence (on log scale) this suggests that the
distribution of $E$ is very narrow\footnote{The ratio of $E$ and the
observed isotropic $\gamma$-ray fluence is a constant of order
100, almost independent of $E$, in the internal shock scenario when
shells are uniform.}  and the large width of the $\gamma$-ray
fluence distribution arises as a result of angular fluctuation in the
$\gamma$-ray emitting surface.

\section{Conclusions and Predictions}
\medskip

A relativistic shell ejected in an explosion could have
large angular fluctuation because regions on the shell
separated by an angle greater than $\gamma^{-1}$ are causally
disconnected. We have explored some consequences of the angular 
fluctuations in GRB explosions. Most of the results described here
also apply to the model consisting of internal collisions
of uniform shells.

We have modeled shells as consisting of independent 
blobs of angular size $\gamma^{-1}$. Because of relativistic beaming 
only a small patch of a shell, of angular size $\gamma^{-1} \sim$  
the size of a blob, is visible to a distant observer. 

The blobs are ejected in the explosion with some distribution of 
Lorentz factor and energy. 
We have calculated the spectrum of emergent photons which are
produced by synchrotron plus inverse Compton processes in internal 
shocks when blobs undergo collisions, and find that the distribution 
of the observed $\gamma$-ray fluence along different lines of sights
is very broad (the total energy in the explosion is fixed at 10$^{52}$ erg).
The width of the fluence distribution
function depends on the width of the distribution of energy
and the Lorentz factor of blobs; the FWHM of the log$_{10}$(fluence) 
distribution function is 1.2 when the distribution of blob energy
is a delta function, where as the width is 2 when the energy 
distribution of blobs, lognormal, has a FWHM of 1.

An intrinsic spread in the energy release in explosions, not considered here,
will broaden the width of the observed $\gamma$-ray fluence distribution. 
The variation of the total explosion energy from one GRB to another 
will however give rise to fully correlated $\gamma$-ray and the afterglow 
emissions, which is at odds with observations.

In this scenario
the emission surface of the expanding shells consists of bright patches,
of angular width of order $\gamma^{-1}$, and dark patches; 
A bright GRB results not because of larger energy release in the
explosion but instead when our line of sight intersects
a bright spot on the expanding colliding shell, and thus the 
$\gamma$-ray fluence is not a good measure of the total energy release
in the explosion.

The fluence distribution function resulting from collisions of uniform
shells is similar in shape but smaller in width by factor of $\sim 10$,
for fixed total energy in explosion, compared to when shells have large
angular fluctuations. For the sub-class of GRBs which show short 
timescale variability, the observed $\gamma$-ray fluence for the 
uniform shell model is proportional to the energy in explosion, and 
the width of the distribution function in this case is equal to the 
width of the total energy release in explosions.

The $\sim 1\%$ radiative efficiency of internal shocks (Kumar, 1999;
Panaitescu, Spada and M\'esz\'aros 1999) in 10--10$^3$ keV energy band
requires total energy in explosion to be larger than the observed
energy in $\gamma$-ray photons by a factor of about 100. The finite
opening angle for burst ejecta reduces the energy requirement by a
factor of 10--100. The angular inhomogeneity of shells ejected in the
explosion could further reduce the energy budget of the brightest
bursts, such as GRB990123, by a factor of $\sim 10$ thereby bringing 
down the total energy involved in the brightest observed bursts to
a value of order the energy in weaker bursts i.e. $\lta 10^{53}$ erg.

An interesting result of this model is that in spite of the very wide
observed luminosity function of GRB, the total energy in GRB explosions
could be roughly comparable in all bursts. This could have interesting
implications on the nature of the inner engine.

There are several predictions of our model. First, the width of the
distribution function in the X-ray afterglow flux should be significantly 
smaller than the spread of fluence seen in $\gamma$-rays. Moreover,
the dispersion of optical luminosity should be smaller than the
X-ray luminosity, and the late time radio afterglow should have the
smallest dispersion which reflects the variation of energy in GRB explosions.
The $\gamma$-ray, X-ray, and the optical data for GRBs with known
redshifts are consistent with these expectations.

It should be noted that the
observed decrease in the width of the isotropic luminosity
distribution for the afterglow emissions, compared to the width of the
isotropic $\gamma$-ray fluence distribution, is contrary to what is
expected if the width of the fluence distribution were a consequence
of a wide distribution of energy release in GRBs (see \S3).

A second prediction is that the afterglow flux should show small 
amplitude fluctuation with time if the energy distribution of blobs 
is not a delta function or shells are not uniform. 
For instance, the fluctuation amplitude 1 day after the explosion,
when the FWHM of blob energy distribution is 10, is $\sim 0.02$ mag
in the optical band and the characteristic variability timescale is
$\sim 1$ day. The early afterglow light curve should show however
larger fluctuations whose amplitude decreases in time.
This prediction could be directly
tested with the forthcoming quick response GRB missions,
HETE II, Swift and BALERINA, and thereby distinguish between the
uniform and fluctuating shell models.

A third prediction is the existence of numerous weak bursts - which
will arise from the low energy tail of the GRB luminosity function.
The afterglow flux from these week bursts should be comparable to the
afterglow from stronger bursts - this may have implications to the
rate of ``orphan'' afterglows.

A fourth prediction is that the fluence distribution of multi peaked
bursts (which arise due to numerous collisions) would be narrower
than the fluence distribution of bursts with only a few peaks. Most of
the bursts detected by BeppoSAX, for which afterglow emission and
redshifts have been measured, show light curves consisting of a few
peaks which could arise as a result of collision of just a few shells.
For such bursts we expect very wide fluence distribution as observed.

Finally, as the prompt optical and the prompt X-ray emissions
arise in regions which are moving with very high Lorentz factor
(Sari and Piran, 1999) we expect these emissions to also have a very wide
luminosity function, whose width should be comparable to the GRB luminosity
function i.e. the prompt emission could be dominated by small hot spots
and produce unusually large fluences in some cases. 
As mentioned earlier we expect temporal fluctuations with a decreasing
amplitude in time during this stage. 

We thank Reem Sari for helpful remarks. 
This research was supported by the US-Israel BSF 95-328, by a grant
from the Israeli Space Agency and by a NASA grant NAG5-3091. TP thanks
F.-K. Thielemann and the Physics department of Basel University for
hospitality while this research was done. PK thanks Savannah for 
encouraging him to investigate the fireball model.


\vfill\eject

\begin{center}
\begin{table*}[ht!]
\begin{tabular}{|c|c|c|c|c|c|c|c|c|}
\hline GRB & Z & F$_\gamma\ ^1$ & F$_\gamma\ ^2$ & E$_{iso}$
&F$_x$(5hr)$\ ^3$ & L$_x$(5hr)$\ ^4$ & R(24hr)& L$_{opt}$(24hr)$\ ^5$ \\
\hline & & $10^{-5}{{\rm ergs\over cm^2}}$ & $10^{-5}{{\rm ergs\over
cm^2}}$ & $10^{53}$ ergs & $10^{-12}{{\rm ergs\over cm^2 sec}}$ &
$10^{46}{{\rm ergs\over sec}}$& & $10^{44}{{\rm ergs\over sec}}$ \\
\hline \hline
990510 & 1.619 & 2.26& 2.94 & 1.5 & - & - & 19.5 &2.8 \\ \hline 990123
& 1.6 & 26.8& 34.9 & 18 & 13.5 & 8.9 & 20.5 & 1.1 \\ \hline 980703 &
.967 & 2.26 & 3.0 & 0.54 & 5.9 & 1.6 & 20.7 & 0.37 \\ \hline 980613 &
1.096 & - & - & - & 0.82 & 2.7 & 22.9 & 0.06 \\ \hline 971214 & 3.412
& 0.944& 1.11 & 3.2 & 1.97 & 4.2 & 22.5 & 0.54 \\ \hline 970828 &
0.958 & 9.60& 13.9 & 3.0 & 5.57 & 1.5 & - & - \\ \hline 970508 & 0.835
& 0.317& 0.55 & 0.054& 1.10 & 0.22 & 21.2 & 0.18 \\ \hline 970228 &
0.695 & - & - & - & - & - & 21.2& 0.13 \\ \hline
\end{tabular}
\hfill\break
\bigskip 
\caption{A comparison of the afterglow and the $\gamma$-ray luminosities
for GRBs with known redshift. The total isotropic energy in $\gamma$-rays, 
the X-ray luminosity after 5 hours in 2--10 keV, and the optical 
luminosity in the R band after 1 day are calculated assuming 
a flat Universe with $\Omega_\Lambda =0.7$ and $h=60$.
\hfill\break
1. The observed fluence in 10keV--2MeV range (from Band et al. 1999).
\hfill\break
2. The $\gamma$-ray fluence after K correction based on a detailed fit of the
spectrum (from Band et al. 1999). 
\hfill\break
3. X-ray flux, 5 hr after the burst, in 2--10 keV energy band.
\hfill\break
4. X-ray luminosity in 2--10 keV using a K correction corresponding to 
a spectrum $\propto \nu^{-0.75}$.
\hfill\break
5. Using a K correction corresponding to a spectrum $\propto \nu^{-0.75}$.
Note that there is no extinction correction.
}
\end{table*}
\end{center}

\vfill\eject

\bigskip
\begin{figure}
\begin{center}
\plotone{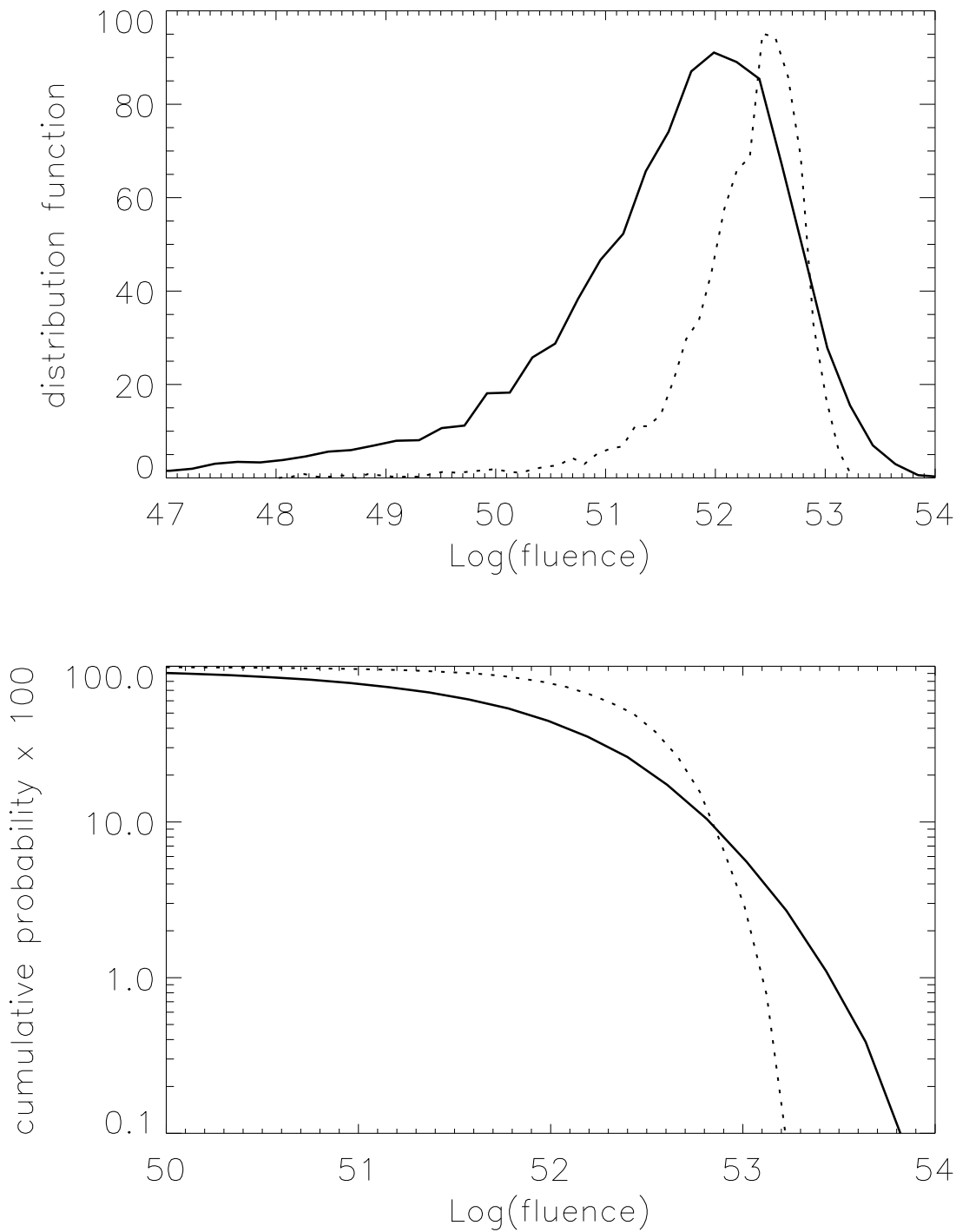}  
\caption{The distribution of GRB fluence in 10--10$^3$ keV band (top
panel).  The mean number of shells ejected along a fixed direction is 5 for 
solid curve and 40 for the dotted curve. The energy in the explosion,
where material is ejected in a cone of opening angle 10$^o$, is taken to 
be 10$^{52}$ erg. Moreover, the burst duration is 30 s, and $\gamma_{min} = 5$ 
\& $\gamma_{max} = 400$, in all cases shown in the figure. The integral 
probability distribution for GRB fluence in 10--10$^3$ keV energy band 
is shown in the bottom panel; the solid and the dotted curves have the
same parameters as the curves in the top panel.}
\end{center}
\end{figure}

\begin{figure}
\begin{center}
\plotone{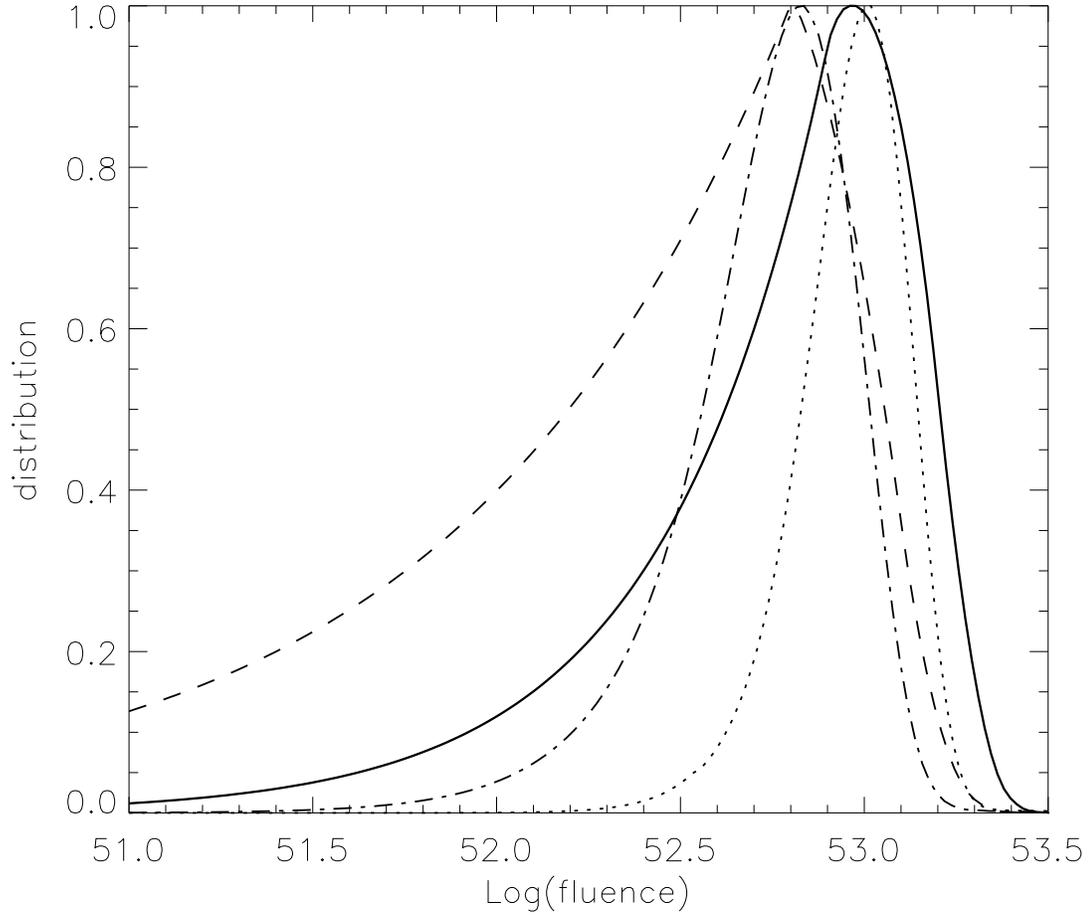} 
\caption{The probability distribution function for the toy model 
discussed in \S2.3. The four different curves are for different values of 
($\alpha, N$) --- (0.5, 4) for the solid curve, (0.5, 10) for the dotted 
curve, and for the dashed and dash-dot curves the values are (0.7, 5) 
and (0.7, 10) respectively. }
\end{center}
\end{figure}
\end{document}